\documentclass[10pt,twocolumn,letterpaper]{article}

\usepackage{cvpr}              

\usepackage{graphicx}
\usepackage{amsmath}
\usepackage{amssymb}
\usepackage{booktabs}
\usepackage{arydshln}

\usepackage{xcolor}

\usepackage[pagebackref,breaklinks,colorlinks]{hyperref}

\usepackage[capitalize]{cleveref}
\crefname{section}{Sec.}{Secs.}
\Crefname{section}{Section}{Sections}
\Crefname{table}{Table}{Tables}
\crefname{table}{Tab.}{Tabs.}

\def\cvprPaperID{*****} 
\def\confName{CVPR}
\def\confYear{2022}

\begin{document}

\title{Continuous-Time Audiovisual Fusion with Recurrence vs. Attention for In-The-Wild Affect Recognition}

\author{Vincent Karas\footnotemark[1]\\
University of Augsburg\\
Augsburg, Germany\\
{\tt\small vincent.karas@informatik.uni-augsburg.de}
\and
Mani Kumar Tellamekala\\
University of Nottingham\\
Nottingham, UK\\
\and 
Adria Mallol-Ragolta\\
University of Augsburg \\
Augsburg, Germany\\
\and
Michel Valstar\\
University of Nottingham\\
Nottingham, UK \\
\and
Björn W. Schuller\\
University of Augsburg\\
Augsburg, Germany\\
}
\maketitle

\begin{abstract}
\noindent
   In this paper, we present our submission to 3rd Affective Behavior Analysis in-the-wild (ABAW) challenge. Learning complex interactions among multimodal sequences is critical to recognise dimensional affect from in-the-wild audiovisual data. Recurrence and attention are the two widely used sequence modelling mechanisms in the literature. To clearly understand the performance differences between recurrent and attention models in audiovisual affect recognition, we present a comprehensive evaluation of  fusion models based on LSTM-RNNs, self-attention and cross-modal attention, trained for valence and arousal estimation. Particularly, we study the impact of some key design choices: the modelling complexity of CNN backbones that provide features to the the temporal models, with and without end-to-end learning. We trained the audiovisual affect recognition models on in-the-wild ABAW corpus by systematically tuning the hyper-parameters involved in the network architecture design and training optimisation. Our extensive evaluation of the audiovisual fusion models shows that LSTM-RNNs can outperform the attention models when coupled with low-complex CNN backbones and trained in an end-to-end fashion, implying that attention models may not necessarily be the optimal choice for continuous-time multimodal emotion recognition. 
   
\end{abstract}


\section{Introduction}
\label{sec:intro}


\noindent
The growing market penetration of \textit{smart} devices is radically increasing the number of scenarios where we interact with machines. Nowadays, such 
interactions take place in a wide range of environments, including the workplace, at home, or even inside our vehicles. If technology is going to accompany us in all aspect of our lives, powering machines with affective capabilities is a requirement to humanise technology towards a more natural \textit{Human-Computer Interaction} (HCI). Creating more human-like technology is one of the objectives of Affective Computing \cite{Picard2010}. 

This paper focuses on the automatic recognition of valence and arousal with the aim to develop Emotional Artificial Intelligence solutions that could allow machines to adapt to the users' affective states. For instance, in the vehicle environment, if the car detects that the driver has been showing high levels of arousal and negative levels of valence, the system could interpret that the driver is experiencing some sort of anger. Alternatively, low levels of arousal and negative valence may indicate sadness or fatigue \cite{Eyben2010}. In this case, the car could suggest playing calm music or even pulling over to take some rest and relax for the safety of the own driver and those in the surroundings. Mood improvement and relaxation systems already exist on the market for some premium brands, but knowing when to suggest them and adapting them based on the detected emotions could greatly enhance the user experience \cite{Braun2019a}.

However, automatically detecting emotions and moods in a setting as described above, or any scenario in an uncontrolled environment, remains an open problem. It is commonly referred to as emotion recognition in the wild, and presents several challenges: Data is often noisy, \eg for the visual modality, a person's face may be partially occluded, or there may be rapid changes in illumination. Audio from the voice may be indistinct 
due to background noise, or missing 
if the person is silent. Another issue lies in cross-cultural emotion recognition \cite{Ringeval2019a}, \ie automatic affect recognition systems needing to perform reliably for people of very diverse backgrounds, who may express their feelings differently. 

In order to tackle this difficult problem, various methods have been proposed. These frequently involve fusing multiple modalities in order to better judge the emotional state from complementary information \cite{Poria2017}. Another common strategy is to make use of temporal information, since the emotional state fluctuates over time. 

For the purpose of processing time series, recurrent neural networks (RNNs) continue to be popular. RNNs look at each element of the input sequentially and update their hidden state. Recent works in emotion recognition have also made use of networks based on self-attention and cross-modal attention \cite{Huang2020,  Chen2020}. While self-attention finds relations between the elements of one sequence, cross-modal attention relates two sequences from different modalities to each other \cite{Tsai2019a}. Compared to RNNs, attention-based network architectures have the advantage of allowing for parallel computation. However, adding recurrence may still improve an attention-based network \cite{Huang2020}. 

Although the recurrence and attention models widely applied to the multimodal fusion for affect recognition and sentiment analysis~\cite{sun2020multi,sun2021multimodal,cai2021multimodal,ma2021hybrid}, it is not very clear how their performance vary depending on the quality of input feature embeddings when modelling complex interactions among the modalities, particularly in end-to-end learning approaches. Specifically, to the best of our knowledge, not much attention is paid to comprehensively analysing the performance of RNNs and attention models based on the underlying CNN backbones’ characteristics. To this end, we consider two commonly used CNN backbone models of two different complexity levels for extracting face image features: FaceNet based on InceptionResNetV1 architecture and MobileFaceNet based on MobileNetV2 architecture. Using the visual features extracted using these two CNN backbones and systematically tuning the  hyper-parameters of network design and optimisation, we comprehensively evaluate the performance of LSTM-RNNs, self-attention and cross-modal attention models on the task of audiovisual affect recognition.  

Herein, we present this comparative analysis of RNNs, self-attention and cross-modal attention as part of our entry to the third Affective Behavior in the Wild (ABAW) challenge. While similar comparisons have been performed, we focus our analysis specifically on the task of continuous emotion recognition in the wild. The Affwild2 dataset used in the challenge is the largest in the wild corpus annotated in terms of valence and arousal \cite{Kollias2019a}. Its data presents many of the difficulties listed above, including a high diversity of subjects, varying illumination and occlusions, and frequently noisy audio or silence. We believe that it is beneficial to benchmark the algorithms on such a dataset.

Our main contributions are:
\begin{enumerate}
    \item We investigate the impact of CNN backbones with different complexities on the performance of LSTM-RNNs for audiovisual affect recognition in the wild, and show the effectiveness of end-to-end-learning.
    \item We contrast the performance of LSTM-RNNs with self-attention and cross-modal attention, and show that recurrent models can outperform attention models in combination with low-complexity CNN backbones.
\end{enumerate}







The rest of the paper is structured as follows: 
We present our methodology in \cref{sec:methodology}, and describe our experimental settings and results in \cref{sec:experiments}. A discussion of the results follows in \cref{sec:discussion}, and make suggestions for future work in \cref{sec:outlook}. Finally, \cref{sec:conclusion} 
concludes this paper.
\section{Related Work}
\label{sec:related}
\noindent \textbf{Recurrence vs. Attention for Sequence Modelling.} To model the underlying temporal dynamics embedded in the continuous-time data, recurrent~\cite{hochreiter1997long} and attention~\cite{Vaswani2017} mechanisms have been widely used. While the recurrence models rely on gated sequential propagation of temporal dynamics encoded into a latent state, the attention models bypass the sequential propagation of information and directly attend to the past inputs. Thus, the attention models can easily capture long-range temporal contingencies by circumventing the problem of vanishing gradients. Although, LSTM-RNNs~\cite{hochreiter1997long} are designed to capture the long-range dependencies by controlling the information flow, they still fall short in practice due to their fixed dimensional latent state to hold the past information, unlike in the attention models. However, this advantage with attention models comes at the cost of poor (quadratic) scalability with the sequence length, which is not the case with RNNs. Furthermore, attention models can operate only within a fixed temporal context window whereas the RNNs can easily handle unbounded context~\cite{gu2021efficiently}. 

Some recent works~\cite{kerg2020untangling, merkx2020human} made systematic efforts to understand the trade-offs between the recurrence and attention mechanisms. However, in the case of continuous-time multimodal fusion which requires modelling complex interactions among different modalities, not much is known about how their performance is influenced by some key design choices, for instance, the CNN backbone modelling complexity and the resultant input features quality. This observation motivates our attempt to study the impact of CNN backbones on the performance of LSTMs, self-attention and cross-modal attention models, by systematically tuning the hyper-parameters involved in the network architecture design and training optimisation. 

\noindent \textbf{In-The-Wild Audiovisual Affect Recognition.}  The first affect in the wild challenge based on the Aff-wild dataset was introduced at CVPR 2017 \cite{zafeiriou2017aff}. In \cite{kollias2019deep}, the dataset and the challenge are described. Aff-wild has 298 videos sourced from YouTube. Shown in it are subjects reacting to a variety of stimuli, \eg film trailers. Subsequently, the corpus was extended with additional videos, and renamed to Aff-wild2 dataset \cite{Kollias2019a}. Aff-wild2 has 548 videos, with a total of about 2,\,78\,M frames. The total number of subjects is 455, 277 of them male. The dataset is annotated with three sets of labels: continuous affect (valence and arousal), basic expressions (six emotions and neutral), and facial action units (FAUs). 545 videos have annotations for valence and arousal.

The first ABAW challenge was held as a workshop at FG2020 \cite{Kollias2020}. It consisted of three sub-challenges for estimating valence-arousal (VA track), classifying facial expressions (EXPR track), and detecting 8 facial action units (AU track). The winning team of the VA track \cite{Deng2020} relied on a  multi-task learning approach. To deal with the problem of incomplete labels in Aff-wild2 data used for the first ABAW competition, \ie not all samples being annotated for each task, Deng et al.~\cite{Deng2020} proposed a teacher-student framework. An ensemble of deep models was trained with semi-supervised learning, where the teacher predicted missing labels to guide the student.

In 2021, the second ABAW challenge took place in conjunction with ICCV 2021 \cite{Kollias2021}. Compared to the previous year, the database had been extended with more annotations. The challenge tracks were identical, but the AU track now included 12 AUs. The winner of the VA track \cite{Deng2021} was the same team as in the previous year, again utilising a multi-task teacher-student framework. The approach also included the prediction uncertainty of an ensemble of student models to further improve performance.


Several multi-task learning models~\cite{Deng2020, Kuhnke2020, do2020affective} effectively leveraged the availability of Aff-wild2 data jointly annotated with the labels of dimensional affect, categorical expressions, and AUs. A holistic multi-task, multi-domain network for facial emotion analysis named FaceBehaviorNet was developed on Aff-wild2 and validated in a cross-corpus setting in \cite{kollias2019face, kollias2021affect, kollias2021distribution}. 

Building on the success of attention mechanism~\cite{Vaswani2017} in sequence data modelling in recent years, cross-modal attention based audiovisual fusion has been widely applied to the emotion recognition tasks~\cite{zhang2020m, fu2021cross,  zhang2021continuous, krishna2020multimodal, rajan2022cross}. Unlike the aforementioned works that solely rely on the fusion of facial and vocal expressions for affect recognition, Antoniadis et al.~\cite{antoniadis2021audiovisual} proposed to use the features of body and background visual context additionally.

\section{Methodology}
\label{sec:methodology}
\begin{figure*}[t]
  \centering
  \includegraphics[width=0.9\textwidth]{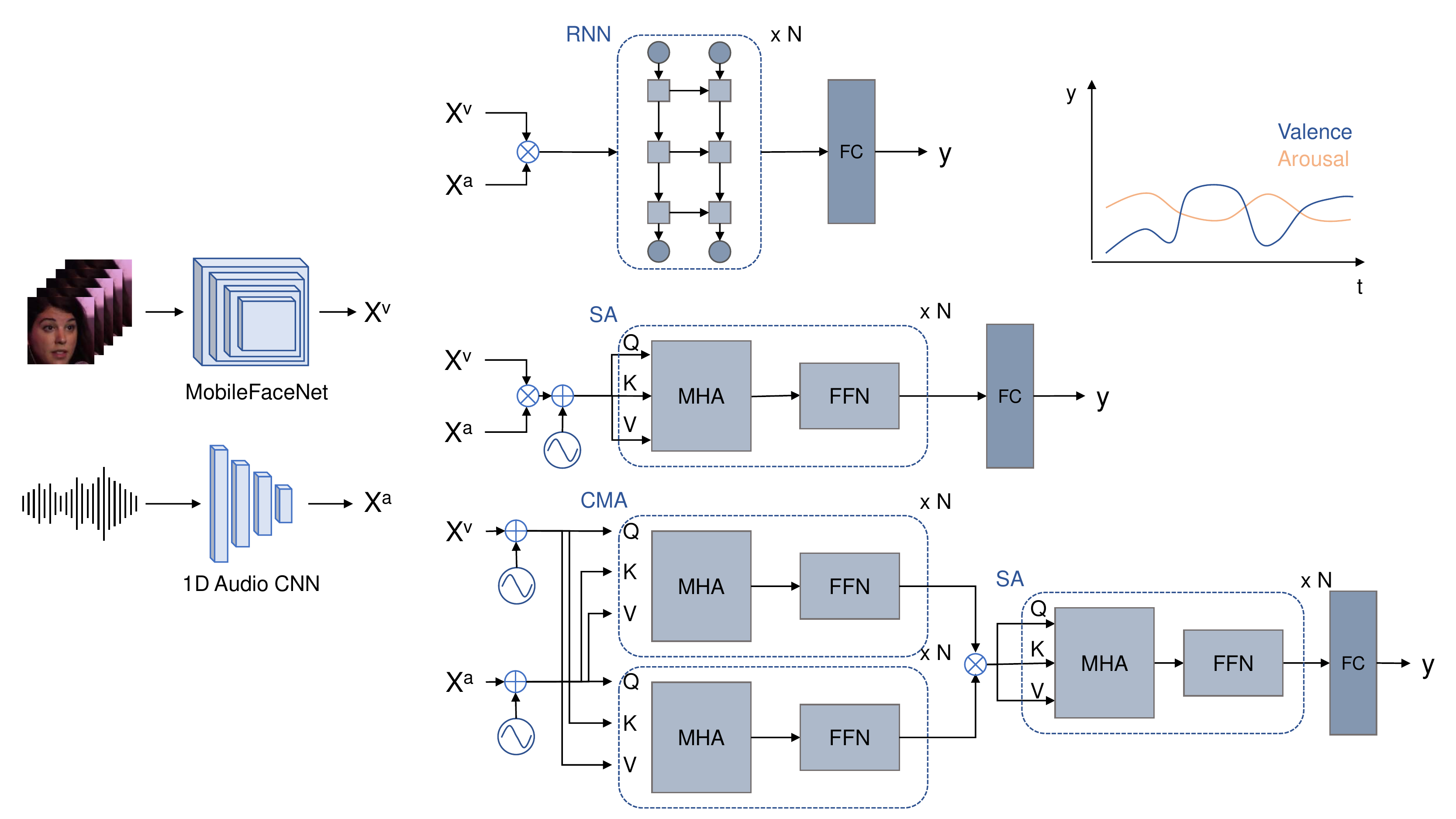}

  \caption{Our proposed approach. We use pre-trained CNNs as feature extractors from sequences faces and raw audio clips (left). Then, we process them with three different architectures: Recurrent Neural Networks with LSTM cells (RNN, top), Self-Attention (SA, middle) and Cross-Modal Attention (CMA, bottom). Each model predicts a sequence of scores for valence and arousal for each timestep of the input (right).}
  \label{fig:approach}
\end{figure*}

\noindent
Since we want to compare fusion methods for time-continuous emotion recognition, our method is based on deep neural networks operating on sequences of features extracted from the visual and audio modalities. We use the cropped and aligned faces from the videos as visual inputs and fixed-length clips as audio inputs. Our approach is illustrated in \cref{fig:approach}.

\subsection{Visual Features}
\noindent
Visual features are extracted with the help of 2D-CNNs. We use pre-trained networks trained on facial recognition tasks. Specifically, we use FaceNet \cite{facenet} based on the InceptionResnetv1 architecture, and trained on VGGFace2 \cite{vggface2}. Alternatively, we employ MobileFaceNet \cite{Chen2018b}, a lightweight architecture designed for facial recognition in embedded devices. MobileFaceNet is built upon residual blocks used in the MobileNetv2 network \cite{Sandler2018} Its usefulness as a feature extractor for emotion recognition was demonstrated in \cite{Deng2021}. Both CNNs return 512-dimensional feature embeddings. The FaceNet has approximately 27M parameters, while the MobileFaceNet has 0.99M parameters.

\subsection{Audio Features}
\noindent
For audio feature extraction, we choose a 1D CNN network based on the architecture proposed by Zhao \etal \cite{Zhao2019a}. The CNN encoder has 4 local feature learning blocks consisting of 1D convolutions and maxpooling layers. The kernel sizes and output channels are [3,3,3,3] and [64, 64, 128, 128]. The choice of this architecture is motivated by its low parameter count (about 88k) and proven effectiveness for speech emotion recognition on a number of corpora.

We use the RECOLA dataset \cite{Ringeval2013}, a corpus of spontaneous affective interactions between French speakers, to pre-train the audio network. For this purpose, we combine the 1D-CNN with a 2-layer LSTM and a fully connected output layer and train the model end-to-end using the  the End2You toolkit \footnote{https://github.com/end2you/end2you} \cite{tzirakis2018end2you}. Then, the LSTM and output layers were removed to obtain the convolutional feature extractor. We then added a global average pooling layer at the end so the network returns 128-dimensional embeddings.

\subsection{Sequence Modelling}

\noindent \textbf{Recurrence Models} are widely used for sequential data modelling, whose fundamental strength lies in their ability to learn the underlying temporal context in the form of a hidden state i.e. $h_t = f(h_{t-1}, ..)$. This approach based on maintaining the hidden states is a natural solution to model the sequential data that is irregularly sampled from an underlying continuous-time series phenomenon~\cite{gu2021efficiently} such as dimensional affect recognition. However, the limitations of recurrence models in terms of capturing cross-modal interactions in multimodal temporal data, which is critical for audiovisual emotion recognition, is not very clear. In this work, we consider the canonical Long-Short Term Memory (LSTM) RNNs~\cite{hochreiter1997long}, using both unidirectional and bidirectional models, for a comprehensive evaluation on valence and arousal estimation from face and speech data.


\noindent \textbf{Self-Attention (SA).} 
Second, we use networks based on the Transformer architecture \cite{Vaswani2017}. Specifically, we use multi-headed scaled dot-product attention blocks with feedforward networks as employed in the transformer encoder. The scaled dot-product attention is defined as:

\begin{equation}
  \text{Attention}(Q,K,V) = \text{softmax}\left(\frac{QK^T}{\sqrt{d_k}}\right)V
  \label{eq:Attention}
\end{equation}

Multi-head attention linearly projects the query, key and value pairs into different sub-spaces and performs attention on them in parallel, before recombining and projecting into the output dimension. It is defined as: 

\begin{equation}
\begin{aligned}
     &\text{MHA}(Q,K,V) = \text{Concat}\left(\text{head}_1, ..., \text{head}_n\right)W^O \\
    & \text{where head}_i = \text{Attention}\left(QW_{i}^{Q}, KW_{i}^K, VW_{i}^{V}\right)
\end{aligned}
\label{eq:MHA}
\end{equation}

In order to fuse modalities within our models, we either use a simple concatenation of our feature embeddings, or a cross-modal fusion architecture.

\noindent \textbf{Cross-Modal Attention (CMA) Fusion} is proposed in Tsai et al.~\cite{Tsai2019a} to implement the Multimodal Transformer network in which pair-wise attention modelling across different modalities is performed. On the task of discrete emotion recognition from multimodal signals, CMA demonstrate superior generalisation performance compared to LSTM-RNNs~\cite{Tsai2019a}. However, when it comes to the continuous emotion recognition from multimodal data, the performance gains that CMA can achieve over the canonical RNNs is unclear. To delineate the trade-offs between the CMA and the other aforementioned sequence models, in this work we evaluate different CMA-based audiovisual fusion models. We implemented audiovisual CMA models by tailoring the multimodal transformer architecture\footnote{ https://github.com/yaohungt/Multimodal-Transformer} which was originally designed for text, audio and visual modalities. 

Our cross-modal architecture is based on the cross-modal attention blocks introduced by \cite{Tsai2019a}. In self-attention used in the transformer encoder, $Q$, $K$ and $V$ are identical. In the cross-modal attention however, the queries and the key, value pairs come from two different modalities, where $Q$ is denoted as the target and $K$,$V$ as the source respectively. It is similar to the transformer decoder, but does not involve self-attention. At each layer, the target modality is reinforced with the low-level information of the source modality \cite{Tsai2019a}.

When employing concatenation of feature vectors, we pass the result through either a stack of recurrent layers  or a self-attention stack. When using cross-modal fusion, we pass the features through two cross-modal blocks in parallel, one of them using the audio features to attend to the visual features and the other vice versa. We then concatenate the outputs of the cross-modal blocks before passing them to a self-attention stack. 

We use fully connected and 1D convolutional layers to reduce the dimension features returned by our extractor networks before passing them to our sequence models. When using 1D-CNNs with kernel size larger than 1, this also serves to encode the local temporal context. For the transformer networks, we add additional position embedding layers with fixed sinusoidal patterns, since they would otherwise not be able to distinguish the order of the sequence passed to them \cite{Vaswani2017}.

\subsection{Loss Functions}
\noindent
We use fully connected layers to return the outputs of our. Each model has two output heads. The first head has size 2 and is used for prediction of valence and arousal scores. 

We use two losses for the regression head. The first is based on the concordance correlation coefficient (CCC) \cite{Lin1989}, which is defined as in \cref{eq:ccc}. It measures the correlation between two sequences, and ranges between -1 and 1, where -1 means perfect anti-correlation, 0 means no correlation, and 1 means perfect correlation. The loss is calculated as $1-CCC$. 

\begin{equation}
\begin{aligned}
    & CCC(x, y) = \frac{2 * cov(x,y)}{\sigma_{x}^2 + \sigma_{y}^2 + \left(\mu_x - \mu_y\right)^2} \\
    & \text{where}\, cov(x,y) = \sum \left(x - \mu_x\right) * \left(y - \mu_y\right)
\end{aligned}
\label{eq:ccc}
\end{equation}

We also compute the mean square error (MSE), which is defined as \cref{eq:mse}. The reasoning behind adding an additional regression loss is that CCC loss alone proved to be less stable during training in our experiments.

\begin{equation}
    MSE(x,y) = \sum \left(x - y\right)^2
\label{eq:mse}
\end{equation}

In addition to regressing the scores, we also add a classification head that predicts the category the scores belong to. Jointly estimating continuous and categorical emotions from faces has been shown to be effective for facial affect analysis in the wild \cite{Toisoul2021}. While the Affwild2 dataset is annotated in terms of both continuous and categorical emotions, the rules of the ABAW challenge do not allow using multiple annotations for the valence-arousal track. Therefore, we discretise the labels, by dividing the two-dimensional affect space into 24 sections. These are derived by plotting valence and arousal in polar coordinates, with 3 equidistant radial subdivisions and 8 angular subdivisions.

Crossentropy loss is used as loss function for the classification head. Since the Affwild2 dataset is imbalanced towards positive arousal and valence, we weigh the logits to emphasise minority classes.

Our total loss is thus composed of three terms. We add weights to the MSE and crossentropy losses to adjust their contribution, leading to our loss function \cref{eq:loss}

\begin{equation}
    \mathcal{L} = \mathcal{L}_{ccc} + \lambda_{mse} * \mathcal{L}_{mse} + \lambda_{ce} * \mathcal{L}_{ce}
\label{eq:loss}
\end{equation}

\section{Experiments and Results}
\label{sec:experiments}
\noindent
We describe or experimental settings and the obtained results on the validation set of the challenge.

\subsection{Dataset}
\noindent
We use subset of Affwild2 annotated for the Valence-Arousal (VA) Estimation task. The training set consists of 341 videos, the validation set consists of 71 videos and the test set consists of 152 videos. Several videos have more than one person in them, those videos are annotated separately for each person and are considered like multiple videos. Frames are annotated with valence and arousal in the range [-1, 1]. 

We use the cropped and aligned faces from the videos provided by the challenge organisers. Some frames are annotated as invalid, after discarding them, we create sequences from the remaining frames. We use a fixed sequence length of 16 frames for our experiments. Audio clips are extracted at a fixed window length of 0.5s, centered at the frame timestamps. We convert the audio of the entire dataset to 16\,kHz mono, 16\,bit PCM.

The frame rate of the Affwild2 dataset is 30fps for the majority of videos. Thus, consecutive frames are very similar. In order to provide our model with more temporal information, one option would be to increase sequence length, at the cost of additional computational resources. We choose instead an approach similar to \cite{Kuhnke2020} and use dilated sampling, \ie we select only 1 in N frames. With sequence length $T$, this gives a temporal context $t$ of:

\begin{equation}
    t = \frac{N}{30} * T
\end{equation}

In order to not reduce the size of the training set, we also apply an interleaved sampling method to select the remaining frames.

We do not apply this dilated sampling method for the validation set. While this introduces some discrepancy with the training, it maintains equal conditions to the test set.

The images are resized to the the shape required by the CNN feature extractor. We use randomly affine transformations and changes in saturation, brightness and contrast as data augmentation on the images. We also apply gaussian noise to the audio frames.

\subsection{Training}
\noindent
We implement out models in the PyTorch framework and train them on servers with Nvidia RTX3090 and A40 GPUs. Per model training, we allocate 40 CPUs and 40GB RAM in order to accelerate the loading of batches. The batch size is 64.

Models are trained using the AdamW optimiser \cite{loshchilov2017decoupled}. We apply cosine annealing with warm restarts as learning rate scheduling, setting it to restart after 200 steps.

In order to find the best configurations for our models, we perform extensive hyperparameter optimisation. We train our models in groups, choosing first the feature extractors and the general architecture (recurrent or transformer), then varying the architecture's parameters as well as the learning rate for our optimiser and the contributions of our losses. A listing of the hyperparameters used is given in.

\begin{table}[h]
    \centering
    \renewcommand{\arraystretch}{1.2} 
    \begin{tabular}{l c}
        \toprule
       \textbf{Hyper-parameter}  & \textbf{Value Range} \\
       \midrule
       \multicolumn{2}{c}{\textbf{General parameters}} \\
       \hdashline
       $n_{layers}$ & [1, 5] \\
       $d_{model}$ & 64, 128, 256 \\
       activation & GELU, SELU \\
       dropout & [0.1, 0.6] \\
       learning rate & [$10^{-5}$, $10^{-2}$] \\
       weight decay & [$10^{-3}$, $10^{-1}$] \\
       $\lambda_{mse}$ & [0.0, 1.0] \\
       $\lambda_{ce}$ & [0.0, 1.0] \\
       \hdashline
       \multicolumn{2}{c}{\textbf{Attention Models}} \\
       $d_{feedforward}$ & 64, 128, 256 \\
       $n_{heads}$ & 2, 4, 8 \\
       \hdashline
       \multicolumn{2}{c}{\textbf{Cross-modal Attention}} \\
       $n_{layers}^{V\xrightarrow{}A}$ & [1, 5] \\
       $n_{layers}^{A\xrightarrow{}V}$ & [1, 5] \\
       \hdashline
       \multicolumn{2}{c}{\textbf{Recurrent Models}} \\
       Context aggregation & \{unidirectional, bidirectional\} \\
       $n_{layers}$ & [1, 5]\\
       $d_{hidden}$ & 64, 128, 256 \\
       \bottomrule
    \end{tabular}
    \caption{Search space of hyper-parameters used for training.}
    \label{tab:my_label}
\end{table}

Since the potential number of hyperparameter combinations is very large, a simple grid search would be inefficient. Instead, we make use of a tuning algorithm to cover a larger number of choices efficiently. For this, we choose Ray Tune \footnote{https://www.ray.io/ray-tune}, a flexible tuning toolkit supports parallel training on multiple GPUs. We use the ASHA scheduling algorithm to quickly discover suitable configurations and stop trials early if they are not performing well.

In a first round of experiments, we freeze the layer weights of the feature extraction networks to limit the number of trainable parameters. Then, we test end-to-end learning with the full set of parameters. For these experiments, we restrict the choice of the visual encoder network to MobileFaceNet to avoid overfitting.

\subsection{Validation Results}
\noindent
The validation results for preliminary experiments on models with frozen feature extraction networks are reported in \cref{tab:my_label}. We denote the three types of architectures employed as Audiovisual-[RNN, SA, CMA] for recurrent, self-attention, and cross-modal attention respectively. The second column specifies the feature extraction network used for visual information,  as Inception or Mobile for InceptionResnetv1 and MobileFaceNet, respectively. For comparison, we also state the results of unimodal models trained with self-attention and RNN.

\begin{table}
    \centering
    \renewcommand{\arraystretch}{1.3} 
    \label{tab:validation_results}
    \begin{tabular}{l l c c c}
    \toprule
        \textbf{Method} & \textbf{Visual CNN} & \textbf{Vale.}  & \textbf{Arou.} & \textbf{Avg.}\\
        \midrule
        Baseline~\cite{kollias2022abaw} & ResNet50 & 0.310 & 0.170 & 0.24 \\
        \hdashline
        \multicolumn{4}{c}{\textbf{Recurrent Models (RNNs)}} \\
         Aud-RNN & - & 0.094 & 0.233 & 0.163 \\
         Vis-RNN & Inception & 0.277 & 0.188 & 0.233 \\
         Vis-RNN & Mobile & 0.285 & 0.357 & 0.321 \\
         AV-RNN & Inception & 0.339 & 0.486 & \textbf{0.413}\\
         AV-RNN & Mobile & 0.319 & 0.436 & 0.378 \\
         \hdashline
         \multicolumn{4}{c}{\textbf{Self-Attention (SA) Models}} \\
         
         Aud-SA & - & 0.076 & 0.317 & 0.197 \\
         Vis-SA & Inception & 0.318 & 0.203 & 0.261 \\
         Vis-SA & Mobile & 0.324 & 0.414 & 0.369 \\
         AV-SA & Inception & 0.344 & 0.404 & 0.374 \\
         AV-SA & Mobile & 0.248 & \textbf{0.529} & 0.389\\
         \hdashline
         \multicolumn{4}{c}{\textbf{Cross-Modal Attention (CMA) Models}} \\
         
         AV-CMA & Inception & \textbf{0.393} & 0.363 & 0.378\\
         AV-CMA & Mobile & 0.324 & 0.460 & 0.392 \\
    \bottomrule
    \end{tabular}
     \caption{The validation results in CCC~$\uparrow$, evaluated on the validation set of Affwild2 as partitioned in ABAW 2022 for unimodal and multimodal models with frozen feature extractors. Valence, arousal and their average are stated for comparison. Results are given for each type of architecture investigated - recurrent network, self-attention and cross-modal attention.}
\end{table}

It can be seen from  \cref{tab:validation_results}  that our audiovisual models outperform the challenge baseline by a wide margin. 

We report the validation results of the best models per architecture, trained end-to-end, in \cref{tab:end_to_end}. All models share the same feature encoders, \ie,  MobileFaceNet and the 1D CNN pre-trained on RECOLA. The hyperparameter configurations of the best models are given in \cref{tab:best_hps}.

\begin{table}[]
    \centering
    \renewcommand{\arraystretch}{1.3} 
    \begin{tabular}{l c c c }
    \toprule
    \textbf{Method} & \textbf{Valence} & \textbf{Arousal} & \textbf{Avg.}\\
    \midrule
         E2E-AV-RNN & 0.361 & \textbf{0.551} & \textbf{0.456} \\
         E2E-AV-SA & 0.380 & 0.520 & 0.450 \\
         E2E-AV-CMA & \textbf{0.388} & 0.492 & 0.440 \\
    \bottomrule
    \end{tabular}
    \caption{Validation results in CCC~$\uparrow$, evaluated on the validation set of Affwild2 in ABAW 2022. Reported results are for the best multimodal models trained end-to-end with MobileFaceNet as visual encoder and 1D CNN pretrained on RECOLA as audio encoder, and using RNN, SA and CMA for sequence modelling.}
    \label{tab:end_to_end}
\end{table}

\begin{table}[]
    \centering
    \renewcommand{\arraystretch}{1.2} 
    \begin{tabular}{l c c c}
    \toprule
       \textbf{Hyper-Parameter}  & \multicolumn{3}{c}{\textbf{E2E Models}} \\
          & AV-RNN & AV-SA & AV-CMA \\
          \midrule
          $n_{layers}$ & 1 & 3 & 4 \\
          $d_{model}$ & 64 & 64 & 256 \\
          activation & SELU & GELU & GELU \\
          dropout & 0.5 & 0.5 & 0.6 \\
          learning rate & 0.0002 & 0.002 & 0.0001 \\
          weight decay & 0.023 & 0.008 & 0.06 \\
          $\lambda_{mse}$ & 0.84 & 0.78 & 0.18 \\
          $\lambda_{ce}$ & 0.88 & 0.27 & 0.76 \\
          $d_{feedforward}$ & - & 256 & 256\\
          $n_{heads}$ & - & 8 & 4 \\
          $n_{layers}^{V\xrightarrow{}A}$ & - & - & 3 \\
          $n_{layers}^{A\xrightarrow{}V}$ & - & - & 1 \\
          Context aggregation & uni & - & - \\
          $d_{hidden}$ & 64 & - & - \\
          \bottomrule
    \end{tabular}
    \caption{Hyperparameter configurations for the best performing models. Models are trained end-to-end with recurrent neural network, self-attention, and cross-modal attention networks, respectively.}
    \label{tab:best_hps}
\end{table}

In addition, we report the number of parameters for the best performing audiovisual models to allow for comparison of computational costs.

\begin{table}[]
    \centering
    \renewcommand{\arraystretch}{1.2}
    \begin{tabular}{l c c c}
    \toprule
       \textbf{Method}  & \textbf{Visual Encoder} & \textbf{P}$_{sequence}$ &  \textbf{P}$_{total}$ \\ 
    \midrule 
    \multicolumn{4}{c}{\textbf{Recurrent Models (RNNs)}} \\
       AV-RNN & Inception & 109\,K & 28.8\,M \\
       AV-RNN & Mobile & 4.4\,M & 5.4\,M \\
       E2E-AV-RNN & Mobile & 76\,K & 1.1\,M \\
    \multicolumn{4}{c}{\textbf{Self-Attention (SA) Models}} \\   
       AV-SA & Inception & 765\,K & 28.1\,M \\
       AV-SA & Mobile & 482\,K & 1.51\,M \\
       E2E-AV-SA & Mobile & 193\,K & 1.2\,M \\
    \multicolumn{4}{c}{\textbf{Cross-Modal Attention (CMA) Models}} \\
       AV-CMA & Inception & 134\,K & 28.1\,M \\
       AV-CMA & Mobile & 2.1\,M & 3.1\,M \\
       E2E-AV-CMA & Mobile & 2.4\,M & 3.4\,M \\
    \bottomrule
    \end{tabular}
    \caption{Size of our models. Shown are the total number of parameters for the audiovisual models, grouped by architecture. For clarity, we report the number of parameters in the sequence models and the full number of parameters separately.}
    \label{tab:parameters}
\end{table}



\section{Discussion}
\label{sec:discussion}
\noindent
When judging performance, we analyse the mean value of CCC for valence and arousal, which is the metric used in the VA Track of the ABAW 2022 challenge. We first discuss how the choice of the visual CNN impacts the RNN models, and the impact of end-to-end learning. We then compare the performance of self-attention and cross-attention, before contrasting RNNs and attention models.

\subsection{RNN performance} \label{ssec:rnn_perf}
\noindent
When using RNN as the sequence model, performance decreases significantly when replacing the FaceNet feature encoder with the less complex MobileFaceNet (0.413 to 0.378). At the same time, the number of parameters in the trainable part of the model increases sharply from 109\,k to 4.4\,M. We interpret this as the model having difficulty to learn valence and arousal effectively from the features returned by the smaller CNN.

However, using the lightweight architecture together with end-to-end-learning presents a very different picture. When the feature extractors are fully trainable, the performance of the recurrent model was greatly increased, yielding an average CCC of 0.456. At the same time, the number of model parameters decreased, to merely 76\,k in the sequence part. Examining the hyperparameter configuration of this winning model showed that it has a single, unidirectional LSTM layer, with a hidden dimension $d=64$.  We conclude from this that the RNN architecture is very efficient in learning representations of the emotional state if it is trained end-to-end in combination with shallow CNN encoders. 

\subsection{Comparing self-attention and cross-attention}\label{ssec:attn_perf}
\noindent
When comparing the self-attention (SA) models with different visual CNNs, it can be seen that the model with MobileFaceNet performs better than the one with FaceNet, with average CCC scores of 0.389 and 0.374, respectively. At the same time, the size of the attention model is smaller for the architecture with MobileFaceNet (482\,k parameters compared to 765\,k parameters). 

For the cross-modal attention (CMA) models, replacing FaceNet with MobileFaceNet also increases performance, from 0.378 to 0.392. However, the transformer network becomes much larger, going from 134\,k parameters to 2.1\,M parameters. As can be seen from these scores, the performance of self-attention and cross-attention appears to be similar if the feature extractors are frozen, with CMA performing slightly better. 

When using end-to-end learning, performance increases significantly for the self-attention model, with an average validation CCC of 0.450. At the same time, the number of parameters in the sequence part of the network shrinks to 193\,k.
The cross-attention model also benefits greatly from end-to-end learning, achieving a score of 0.44. The number of parameters in the sequence part of the model is 2.4\,M.

Comparing the two best models trained end-to-end shows that the self-attention model outperforms the cross-attention, while being significantly smaller. We hypothesise that the lower complexity of the self-attention model helped discover a more efficient architecture during end-to-end training. 

\subsection{Comparison between RNN and attention for sequence modeling}
\noindent
We now compare the performances of our RNN models and attention models directly based on the results discussed in the two previous sections. In the case the feature extractors are frozen, for the larger FaceNet, the RNN outperforms the attention models. If frozen MobileFaceNet is used, the attention models outperform RNN. With end-to-end learning, RNN beats both self-attention and cross-attention, while also having fewer parameters.

We conclude from this that our initial hypothesis that attention-based models consistently outperform RNNs for emotion recognition in the wild has not been confirmed. When end-to-end learning is used in combination with shallow CNNs for feature encoding, RNNs perform superior to the attention-based models investigated in this paper. 



\section{Outlook}
\label{sec:outlook}
\noindent
We compared fusion performance using two CNNs of different sizes as visual feature extractors, while using a small 1D-CNN for extracting audio features. Another study could focus on choosing different audio networks, \eg a larger model like VGGish \cite{Hershey2017}, and comparing the effects.

The models used in this work had limited temporal context due to computational constraints. Future studies could extend towards greater sequence lengths to investigate how well the models capture long-term dependencies.

Our analysis has focused on the average of valence and arousal as a metric, in order to judge the overall performance of the models. We leave the analysis of trade-offs between valence and arousal for future work.

\section{Conclusion}
\label{sec:conclusion}
\noindent
On a wide range of sequence modelling tasks, attention models demonstrated superior generalisation performance than recurrent models in recent years. However, it is worth noting that the recurrent models have the natural ability to cope with the challenges in learning from time-continuous sequence data, by inferring the latent states with unbounded context, at least in principle. Therefore, in the case of continuous-time multimodal affect recognition, a recurrent neural network architecture may still be a natural choice to model the latent states of face and voice data and their interactions in a time-continuous manner. Extensive evaluation of LSTM-RNNs, self-attention and cross-modal attention on in-the-wild audiovisual affect recognition, suggests that attention models may not necessarily be the optimal choice to perform continuous-time multimodal information fusion. 

{\small
\bibliographystyle{ieee_fullname}
\bibliography{egbib}
}

\end{document}